\documentclass[aps,preprint]{revtex4}%
\usepackage{amsfonts}
\usepackage{amsmath}
\usepackage{amssymb}
\usepackage{graphicx}%
\newtheorem{theorem}{Theorem}

\newtheorem{lemma}{Lemma}

\newenvironment{proof}[1][Proof]{\noindent\textbf{#1.} }{\ \rule{0.5em}{0.5em}}
\begin{document}
\preprint{Internal Communication }
\title{Cosmic microwave background dipole, peculiar velocity and Hubble flow}
\author{Yukio Tomozawa}
\affiliation{Michigan Center for Theoretical Physics}
\affiliation{Randall Laboratory of Physics, University of Michigan}
\affiliation{Ann Arbor, MI. 48109-1040, USA}
\date{\today }

\begin{abstract}
Two types of cosmology are discussed and their implications for the observed
cmb (cosmic microwave background radiation) dipole are described.\ Theorems
useful for understanding the cause for a cmb dipole are presented. Since the
present peculiar velocity of the solar system relative to the GA (Great
Attracter) cannot explain the observed cmb dipole, the author presents the
possibility of Hubble flow of the GA as a cause in one case and a further
peculiar velocity of the GA in the other case.

\end{abstract}

\maketitle

\section{\label{sec:level1}Introduction}

There are two types of cosmology even in the realm of the Friedman universe,
one with the center for the expansion of the universe and the other without a
center. The Hubble law, \textbf{v }= H$_{\text{0}}$ \textbf{r}, yields the
relationship, \textbf{v}$_{\text{2}}$ - \textbf{v}$_{\text{1}}$= H$_{\text{0}%
}$ ( \textbf{r}$_{\text{2}}$ - \textbf{r}$_{\text{1}}$) for any two galaxies
with positions and velocities, \textbf{r}$_{\text{1}}$, \textbf{v}$_{\text{1}%
}$ and \textbf{r}$_{\text{2}}$ , \textbf{v}$_{\text{2}}$ respectively, where
H$_{\text{0}}$ = 100 h km/s-Mpc is the Hubble constant (with h = 0.5
\symbol{126}0.85). For convenience of discussion, we assume the value of
H$_{\text{0}}$ to be 70.0 km/s-Mpc in this article. This equation implies that
every point appears to be the center of the expansion. In other words, both
types of cosmology yield the same conclusion, as far as the Hubble law is
concerned. However, the observed cmb dipole has different implications for the
two types of cosmology and observational differences are discussed.

\section{Two Types of Cosmology}

In the Friedman universe,%
\begin{equation}
ds^{2}=dt^{2}-a(t)^{2}(dr^{2}/(1-kr^{2})+r^{2}d\theta^{2}+r^{2}\sin^{2}%
(\theta)d\phi^{2}),
\end{equation}
with an appropriate source, $T_{\nu}^{\mu}$ , there are two types of
interpretation for the radial coordinate, $r$.

I) Cosmology A

The origin of the radial coordinate has a physical meaning as a point where
the expansion started. Each point of the universe has a Hubble flow velocity
relative to the origin that is proportional to the distance from the origin.
As is discussed in the introduction, the linearity of the Hubble law makes
every point in the universe look like a center for the expansion of the universe.

II) Cosmology B

The universe resides on the surface of an expanding balloon. The center does
not exist in the universe. (It exists outside the universe.) The origin of the
coordinate can be chosen at any point but there is no special significance for
such a choice. The Hubble law is naturally built into the framework. There is
no velocity associated with each point of the universe, but the relative
distance and relative velocity of any two points increase with the expansion
of the balloon. Each point is equivalent relative to a distant cmb emittor and
there is no cmb dipole at any point except that due to a peculiar velocity in
a cluster.

I will discuss the implication of the observed cmb dipole in cosmologies A and B.

\section{The CMB Dipole in the Temperature Distribution in Cosmology A}

A dipole component was observed in the temperature distribution in the cmb
measurement. The cmb dipole for blueshift for the solar system \cite{dipole}
is given by%
\begin{equation}
v=371\pm0.5\text{ }km/s,\text{ \ \ \ }l=264.4\pm0.3%
{{}^\circ}%
,\text{ \ \ \ }b=48.4\pm0.5%
{{}^\circ}%
.
\end{equation}
Or equivalently, the cmb dipole for redshift is%
\begin{equation}
v=371\pm0.5\text{ }km/s,\text{ \ \ \ }l=84.4\pm0.3%
{{}^\circ}%
,\text{ \ \ \ }b=-48.4\pm0.5%
{{}^\circ}%
. \label{dipole}%
\end{equation}
By using the observation of a peculiar velocity for the solar
system\cite{sciama},\cite{ga}, one can compute the cmb dipole component of the
cluster (Virgo) center and that of the supercluster (GA) center. The detailed
calculation of the cmb dipoles at the cluster centers will be performed below,
and it will be shown that the cmb dipoles at the cluster centers are much
larger than that of the solar system. If the cmb dipole at the cluster center
were zero, one could conclude that the observed cmb dipole would be due to the
peculiar velocity. In other words, the observed cmb dipole cannot be explained
in terms of the peculiar velocity. In order to understand the meaning of the
observed cmb dipole, the author presents pertinent theorems.

\begin{theorem}
With the assumption of the existence of a center for expansion of the
universe, Hubble flow creates a cmb dipole with redshift in the direction of
the Hubble flow with the magnitude of the Hubble flow velocity.
\end{theorem}

\begin{proof}
Let the velocities of the Hubble flow and the cmb emitter in the direction of
the Hubble flow be $v_{H}$ and $v$, respectively. Relating an equivalent
velocity of the cmb emittor $v$ to the expansion rate $1+z$ by%
\begin{equation}
\sqrt{\frac{1+v/c}{1-v/c}}=1+z,
\end{equation}
one gets%
\begin{equation}
v/c=\frac{(1+z)^{2}-1}{(1+z)^{2}+1}=1-2\frac{1}{(1+z)^{2}}=1-2x10^{-6}%
\end{equation}
for $z=1000$. The relative velocity is $v_{+}=(v-v_{H})/(1-vv_{H}/c^{2})$. The
analogous velocities in the opposite direction are $-v_{H}$ and $v-2v_{H}$,
and the relative velocity is $v_{-}=(v-2v_{H}+v_{H})/(1+(v-2v_{H})v_{H}%
/c^{2})$ in the opposite direction. These relative velocities are then
expressed for \ $v\approx c$ and $v_{H}\ll c$
\begin{equation}
v_{+}\approx v-v_{H}+v_{H}%
\end{equation}
and%
\begin{equation}
v_{-}\approx v-v_{H}-v_{H}.
\end{equation}
This proves the statement. Obviously, the cmb dipole vanishes at the center of
the universe.
\end{proof}

\begin{theorem}
The cmb dipoles for redshift at the center of a cluster and at a member galaxy
with a peculiar velocity $\mathbf{v}_{p}$ are related by%
\begin{equation}
\mathbf{v}(dipole\text{ }at\text{ }the\text{ }clustercenter)=\mathbf{v(}%
dipole\text{ }at\text{ }the\text{ }galaxy)-\mathbf{v}_{p}+\mathbf{v(}%
clustercenter-galaxy)
\end{equation}
where%
\begin{equation}
\mathbf{v}(A-B)=\mathbf{v}(A)-\mathbf{v}(B)
\end{equation}

\end{theorem}

\begin{proof}
The peculiar velocity plays a role similar to Hubble flow, as far as the cmb
dipole is concerned. This can be seen by considering the case where the
peculiar velocity $\mathbf{v}_{p}$ and Hubble flow $\mathbf{v}_{H}$ are
parallel. In this case, the galaxy is at rest with respect to a galaxy with
Hubble flow, $\mathbf{v}_{H}+\mathbf{v}_{p}$. Since both galaxies should have
the same cmb dipole, the above statement is established. The term,
$\mathbf{v}(A)-\mathbf{v}(B)=H_{0}(\mathbf{r}_{A}-\mathbf{r}_{B})$,\ is for
adjustment of the Hubble law. In fact, the necessity of the term,
$\mathbf{v}(center-galaxy)=\mathbf{v}(center)-\mathbf{v}(galaxy)$, is
understood when the formula is applied to galaxies at the clustercenter and a
galaxy in the limit of vanishing peculiar velocity.
\end{proof}

\begin{lemma}
The center of the universe is characterized as a point of the universe where
the cmb dipole has an intrinsically vanishing value.
\end{lemma}

The term intrinsically vanishing value is used, since an accidentally
vanishing value for a cmb dipole can occur for a point galaxy in a cluster for
which the Hubble flow of the cluster and the peculiar velocity of the galaxy
have identical magnitude and opposite direction. Clearly, that galaxy is at
rest relative to the center of the universe at that instant. Hereafter a cmb
dipole implies a redshift dipole unless otherwise stated.

The peculiar velocity of the sun relative to the Virgo center of the local
cluster is estimated to be \cite{sciama}
\begin{equation}
v=415\text{ }km/s,\text{ \ \ \ }l=335%
{{}^\circ}%
,\text{ \ \ \ }b=7%
{{}^\circ}
\label{sciama1}%
\end{equation}
or%
\begin{equation}
v=630\text{ }km/,\text{ \ \ \ }l=330%
{{}^\circ}%
,\text{ \ \ \ }b=45%
{{}^\circ}
\label{sciama2}%
\end{equation}
We examine these two possibilities. The outcomes are listed in this order for
each case. Using Theorem 2, one computes the cmb dipole at the Virgo center,
which is located at%
\begin{equation}
v=1050\pm200\text{ }km/s,\text{ \ \ \ }l=287%
{{}^\circ}%
,\text{ \ \ \ }b=72.3%
{{}^\circ}%
\end{equation}
corresponding to a distance of $15\pm3$ Mpc. The key formula is%
\begin{equation}
\mathbf{v}(dipole\text{ }at\text{ }Virgo)=\mathbf{v}(dipole\text{ }at\text{
}the\text{ }Sun)-\mathbf{v}_{p}(Sun/Virgo)+\mathbf{v}(Virgo) \label{virgo}%
\end{equation}
For the cmb dipole at the Virgo center, one obtains
\begin{equation}
v=728.3\pm148\text{ }km/s,\text{ \ \ \ }l=336.0\pm8.2%
{{}^\circ}%
,\text{ \ \ \ }b=67.4\pm11.3%
{{}^\circ}%
\end{equation}%
\begin{equation}
v=418.8\pm47km/s,\text{ \ \ \ }l=328.8\pm6.4%
{{}^\circ}%
,\text{ \ \ \ }b=41.5\pm28.0%
{{}^\circ}%
\end{equation}
depending on the two choices of peculiar velocity. We note that the Cartesian
coordinates for $(v,l,b)$ are expressed as $(v\cos b\cos l,v\cos b\sin l,v\sin
b)$.

Further, the Virgo cluster is considered to be part of a supercluster centered
around the GA. In order to compute the cmb dipole at the GA, one assumes the
position of the GA to be \cite{ga}%
\begin{equation}
v=4200\text{ }km/s,\text{ \ \ \ }l=309%
{{}^\circ}%
,\text{ \ \ \ }b=18%
{{}^\circ}
\label{eqga1}%
\end{equation}
or%
\begin{equation}
v=3000\text{ }km/s,\text{ \ \ \ }l=305%
{{}^\circ}%
,\text{ \ \ \ }b=18%
{{}^\circ}
\label{eqga2}%
\end{equation}
and the infall velocity of the Virgo center to the GA to be
\begin{equation}
v_{in}=1000\pm200km/s \label{infallv}%
\end{equation}
The direction of the infall is determined by%
\begin{equation}
\mathbf{v}(GA-V)=\mathbf{v}(GA)-\mathbf{v}(Virgo)
\end{equation}
resulting in%
\begin{equation}
v(GA-V)=3712.3\pm75\text{ }km/s,\text{ \ \ \ }l(GA-V)=310.9\pm0.4%
{{}^\circ}%
,\text{ \ \ \ }b(GA-V)=4.6\pm2.8%
{{}^\circ}%
\text{ \ \ \ } \label{infall1}%
\end{equation}
for Eq.(\ref{eqga1}) and%
\begin{equation}
v(GA-V)=2552.5\pm59\text{ }km/s,\text{ \ \ \ }l(GA-V)=307.2\pm1.6%
{{}^\circ}%
,\text{ \ \ \ }b(GA-V)=1.6\pm3.0%
{{}^\circ}
\label{infall2}%
\end{equation}
for Eq. (\ref{eqga2}).

Using Theorem 2, one can compute the cmb dipole at the GA by%
\begin{equation}
\mathbf{v}(dipole\text{ }at\text{ }GA)=\mathbf{v}(dipole\text{ }at\text{
}Vigo)-\mathbf{v}(Virgo\text{ }infall)+\mathbf{v}(GA-V), \label{ga}%
\end{equation}
where the direction of the infall is given by Eq. (\ref{infall1}) or Eq.
(\ref{infall2}). Then, the cmb dipole at the GA is given by%
\begin{equation}
v=2609.3\pm120\text{ }km/s,\text{ \ \ \ }l=308.1\pm0.2%
{{}^\circ}%
,\text{ \ \ \ }b=19.9\pm3.4%
{{}^\circ}%
\end{equation}%
\begin{equation}
v=2459.1\pm98\text{ }km/s,\text{ \ \ \ }l=308.6\pm0.6%
{{}^\circ}%
,\text{ \ \ \ }b=11.6\pm3.9%
{{}^\circ}%
\end{equation}
for Eq. (\ref{eqga1}), and%
\begin{equation}
v=1455.6\pm142\text{ }km/s,\text{ \ \ \ }l=301.3\pm0.6%
{{}^\circ}%
,\text{ \ \ \ }b=25.5\pm5.3%
{{}^\circ}%
\end{equation}%
\begin{equation}
v=1286.6\pm104\text{ }km/s,\text{ \ \ \ }l=302.0\pm0.7%
{{}^\circ}%
,\text{ \ \ \ }b=10.4\pm7.3%
{{}^\circ}%
\end{equation}
for Eq. (\ref{eqga2}).

Based on Theorem 1, one may assume that the cmb dipole at the GA calculated
above is due to the Hubble flow of the center of the GA supercluster. Then,
one can compute the position of the center for expansion of the universe by%
\begin{equation}
\mathbf{v}(universe\text{ }center)=\mathbf{v}(GA\text{ }center)-\mathbf{v}%
(cmb\text{ }dipole\text{ }at\text{ }GA) \label{center}%
\end{equation}
The position of the universe center thus obtained is%
\begin{equation}
v_{c}=1595.3\pm196\text{ }km/s,\text{ \ \ \ }l_{c}=310.5\pm0.1%
{{}^\circ}%
,\text{ \ \ \ }b_{c}=14.8\pm1.5%
{{}^\circ}%
\end{equation}%
\begin{equation}
v_{c}=1779.6\pm184\text{ }km/s,\text{ \ \ \ }l_{c}=309.7\pm0.2%
{{}^\circ}%
,\text{ \ \ \ }b_{c}=26.8\pm2.7%
{{}^\circ}%
\end{equation}
for Eq. (\ref{eqga1}), and%
\begin{equation}
v_{c}=1585.4\pm196\text{ }km/s,\text{ \ \ \ }l_{c}=308.1\pm0.2%
{{}^\circ}%
,\text{ \ \ \ }b_{c}=13.0\pm1.6%
{{}^\circ}%
\end{equation}%
\begin{equation}
v_{c}=1759.3\pm181\text{ }km/s,\text{ \ \ \ }l_{c}=307.4\pm0.0%
{{}^\circ}%
,\text{ \ \ \ }b_{c}=25.3\pm2.9%
{{}^\circ}%
\end{equation}
for Eq. (\ref{eqga2}). Conversion to the ordinary distance scale yields
$22.8\pm2.8$ $Mpc$, $25.4\pm2.6$ $Mpc$ for Eq. (\ref{eqga1}) and $22.6\pm2.8$
$Mpc$, $25.1\pm2.6$ $Mpc$ for Eq. (\ref{eqga2}).

The last set of expressions for the universe center can be obtained
alternatively directly from the total peculiar velocity of the solar system
towards the GA supercluster center and the cmb dipole of the sun. Using Eq.
(\ref{virgo}), Eq. (\ref{ga}) and Eq. (\ref{center}), one gets%
\begin{equation}
\mathbf{v}(universe\text{ }center)=-(\mathbf{v}(dipole\text{ }at\text{
}the\text{ }Sun)-\mathbf{v}_{p}(total)) \label{center2}%
\end{equation}
where%
\begin{equation}
\mathbf{v}_{p}(total)=\mathbf{v}_{p}(Sun/Virgo)+\mathbf{v}_{p}(Virgo/GA)
\end{equation}
is the total peculiar velocity of the sun towards the GA. From Eq.
(\ref{sciama1}), Eq. (\ref{sciama2}), Eq. (\ref{infallv}), Eq. (\ref{infall1})
and Eq. (\ref{infall2}), one finds the total peculiar velocity to be%
\begin{equation}
v=1389.2\pm198\text{ }km/s,\text{ \ \ \ }l=317.9\pm1.2%
{{}^\circ}%
,\text{ \ \ \ }b=5.4\pm\text{ }0.1%
{{}^\circ}%
\text{\ }%
\end{equation}%
\begin{equation}
v=1519.1\pm190\text{ }km/s,\text{ \ \ \ }l=316.8\pm0.9%
{{}^\circ}%
,\text{ \ \ \ }b=20.2\pm2.4%
{{}^\circ}%
\text{ \ \ \ \ }%
\end{equation}
for Eq. (\ref{eqga1}), and%
\begin{equation}
v=1379.7\pm189\text{ }km/s,\text{ \ \ \ }l=315.2\pm1.4%
{{}^\circ}%
,\text{ \ \ \ }b=3.3\pm\text{ }0.2%
{{}^\circ}%
\text{\ }%
\end{equation}%
\begin{equation}
v=1497.6\pm189km/s,\text{ \ \ \ }l=314.2\pm1.1%
{{}^\circ}%
,\text{ \ \ \ }b=18.4\pm\text{ }2.5%
{{}^\circ}%
\text{\ }%
\end{equation}
for Eq. (\ref{eqga2}). Application of Eq. (\ref{center2}) yields the same
result for the coordinates of the center of the universe obtained above.

In the end, the author has arrived at the following conclusion.

\begin{theorem}
The Hubble flow of the solar system is nothing but%
\begin{equation}
\mathbf{v}_{H}(the\text{ }Sun)=\mathbf{v}(the\text{ }cmb\text{ }dipole\text{
}at\text{ }the\text{ }Sun)-\mathbf{v}_{p}(the\text{ }total\text{ peculiar
}velocity\text{ }of\text{ }the\text{ }Sun). \label{hubble}%
\end{equation}

\end{theorem}

\section{The case for Cosmology B}

In this case, there is no cmb dipole at any point except due to a peculiar
velocity\ in a cluster. Therefore%
\begin{equation}
\mathbf{v(}dipole\text{ }at\text{ }the\text{ }Sun)-\mathbf{v}_{p}%
(total)-\mathbf{v}_{p}(GA/X)=0
\end{equation}
or equivalently%
\begin{equation}
\mathbf{v}_{p}(GA/X)=\mathbf{v(}dipole\text{ }at\text{ }the\text{
}Sun)-\mathbf{v}_{p}(total) \label{pX}%
\end{equation}
must be a peculiar velocity of the GA supercluster towards an unknown center
X. Comparing Eq (\ref{center2}) and Eq (\ref{pX}) or Eq (\ref{hubble}), it is
obvious%
\begin{equation}
\mathbf{v}_{p}(GA/X)=-\mathbf{v}(universe\text{ }center)=\mathbf{v}%
_{H}(the\text{ }Sun).
\end{equation}
In other words, -(the center coordinate) or the Hubble flow of the sun in
cosmology A plays the same role as the extra peculiar velocity of the GA
towards unknown object X.

\section{Summary and Discussion}

The observed cmb dipole cannot be explained by the present status of the
peculiar velocity of the solar system, neither relative to the center of the
Virgo cluster, nor relative to the center of the GA supercluster. Instead, it
implies a large cmb dipole at the location of the GA. As far as I know, up to
now there has been no explanation for such a cmb dipole at a supercluster
center. Using Theorem 1, the author has presented an explanation for a cmb
dipole at the GA in terms of Hubble flow. In other words, the author presented
an explanation for the observed cmb dipole as a result of the Hubble flow of
the GA and the peculiar velocity of the solar system towards the GA. This
interpretation inevitably determines the location of the center for the
expansion of the universe in cosmology A. Alternatively, in cosmology B the GA
cluster center must have a peculiar velocity towards X in the southern
hemisphere. It is important to determine whether a peculiar velocity for the
GA center with the value given by Eq. (\ref{pX}) can be found by future
observation. If not, cosmology A prevails.

\begin{acknowledgments}
The author would like to thank the members of the Physics Department and the
Astronomy Department of the University of Michigan for useful information.
\end{acknowledgments}

Correspondence should be addressed to the author at tomozawa@umich.edu.

\end{document}